\documentclass[final,3p,times,twocolumn]{elsarticle}

\usepackage{graphicx}

\usepackage{amssymb}

\journal{Physica C}

\newcommand{\beq}{\begin{equation}}
\newcommand{\eeq}{\end{equation}}
\newcommand{\beqn}{\begin{eqnarray}}
\newcommand{\eeqn}{\end{eqnarray}}
\newcommand{\bea}{\begin{eqnarray}}
\newcommand{\eea}{\end{eqnarray}}

\begin{document}

\begin{frontmatter}

\title{Nematic orders in Iron-based superconductors}

\author[1,2]{Jiangping Hu}

\author[3]{Cenke Xu}

\address[1]{Department of Physics, Purdue University, West Lafayette,
Indiana 47907, USA}
\address[2]{Beijing National Laboratory for
Condensed Matter Physics, and Institute of Physics, Chinese
Academy of Sciences, Beijing 100190, China}
\address[3]{Department of Physics, University of California, Santa
Barbara, CA 93106}




\begin{abstract}
In the newly discovered iron-based superconductors, many
experiments have demonstrated the existence of the rotational
symmetry breaking nematic order, which has been a prevailing
phenomenon in many correlated electronic systems. In this paper,
we review nematic behaviors in iron-pnictides and the mechanism
behind the development of the nematic order. We discuss evidences
that support the spin-driven nematicity in iron-pnictides.
Theories, results and predictions will be discussed based on this
picture. We also briefly discuss the generalization of this theory
to the nematicity in iron-chalcogenides.

\end{abstract}


\end{frontmatter}







\section{Introduction}

Microscopically, strongly correlated systems are usually described
by extended Hubbard models with parameters such as Hubbard
interaction, Hunds coupling, doping, etc. Despite the relatively
simple microscopic models, the infrared physics of strongly
correlated systems can be incredibly rich, various phases with
completely different low energy descriptions can appear in the
same phase diagram tuned by a few microscopic tuning parameters.
Directly deriving the precise low energy phase diagram from the
microscopic extended Hubbard models is usually tremendously
difficult except for some special limits of the model. However,
the competition and interplay between low energy phases can still
be reasonably well described using effective theories that are
based on some fundamental rules of physics. Many qualitative and
semi-quantitative conclusions can be already drawn based on these
low energy effective theories without the detailed information of
the microscopic physics of the system.

Two decades after the discovery of the Cu-based superconductors
(curpates), a completely new class of high temperature
superconductors was found, where the Fe$^{2+}$ ions play the most
important role\cite{tc01,tc02,tc03,tc04,tc05,tc06,tc07}.
Iron-based superconductors can be divided into two basic
categories: iron-pnictides and iron-chalcogenides. The category of
iron-pnictides includes many different families of materials which
are named as the 1111-family (example: LaOFeAs)
\cite{tc01,tc05,tc07}, the 111-family (example:
$\mathrm{NaFeAs}$), the 122 family(example: $\mathrm{BaFe_2As_2}$)
\cite{Rotter2008}, and more complicated structures such as
$\mathrm{Sr_4V_2O_6Fe_2As_2}$\cite{Zhuxy2009}. Iron-chalcogenides
include the 11 family Fe/Se(Te)\cite{Hsu2008} and the 122
family-alkali doped  $\mathrm{A_{1-x}Fe_{2-y}Se_2 \ (A=K,Cs,Rb)}$
\cite{Guo2011a,Fang2011a, Liu2011a, Yan2011a}. The transition
temperatures of these new superconductors reach as high as
56K\cite{tc07} for iron-pnictides and 48K\cite{sunlili2011} for
iron-chalcogenides.

The discovery of Fe-based superconductors provides a precious
opportunity to investigate possible universalities shared in all
of high temperature superconductors. Fe-based superconductors
share many common features with cuprates. Both of them are
quasi-two dimensional systems with a similar doping-dependent
phase diagram involving strong antiferromagnetism (AF)\cite{La1}.
Superconductivity is developed by introducing doping or applying
pressure while long-range magnetic orders are suppressed (for a
review, see Ref.~\cite{review-johnston2010}). Understanding the
magnetism is thus pivotal to illustrate superconducting mechanism
in these materials. While the relation between magnetism and
superconductivity has been a central research focus in the past
several years, here we review another important aspect of
magnetism: electron nematicity, in undoped or under-doped iron
based superconductors.

An electronic nematic liquid phase is characterized by rotational
symmetry breaking. In the past decade, this phase has been shown
to appear almost universally in quasi-two dimensional electronic
systems. The nematicity can arise from weakly correlated Fermi
liquid systems dominated by kinetic energy to strongly correlated
electron systems where electron-electron interactions are
important. It has been observed throughout a variety of different
electron systems including two-dimensional electron systems with
high magnetic fields, strontium ruthenate materials and cuprates
(for a review, see Ref.~\cite{fradkin-nematic}).

In undoped or under-doped iron-pnictides, transport measurements
reveal the existence of an intrinsic anisotropy of in-plane
resistivity above magnetic Neel transition temperature
$T_N$\cite{nematic-fisher2011, ying2011chen}, a strong evidence
for the presence of an electronic nematic state. Similar
anisotropy appears in a variety of experimental measurements,
including local electronic structures measured by scanning
tunneling microscopy(STM)\cite{nematic-chuang2010}, magnetic
fluctuations by neutron
scattering\cite{Zhaojun2009,Harringer2010}, dynamic conductivity
by optical reflection measurements \cite{Nakajima-nematic-optical}
and electronic structures measured by angle resolved photoemission
spectroscopy (ARPES) \cite{nematic-yi2010,Yi2011-nematic} in
de-twinned samples.

However, it is still controversial whether the observed
anisotropies indicate an indisputable electronic nematic state.
The debate arises because of the interplay between multi-degrees
of freedom and the complexity of electronic structures in
iron-pnictides. The lattice, spin, and orbital degrees of freedom
all manifest themselves in the nematic state, which causes a
debate on the origin of the rotational symmetry breaking of the
tetragonal lattice. The magnetically ordered state in
iron-pnictides is a collinear-AF (C-AF) state whose ordered
wavevectors are $(0,\pi)$ or $(\pi,0)$ with respect to the
tetragonal iron lattice. The C-AF state not only breaks the SO(3)
symmetry of the spin space, but also breaks the $C^4$ rotational
symmetry of the tetragonal lattice. Therefore, it is possible that
an electronic nematic order which only breaks the latter $C^4$
symmetry can be induced by magnetic
fluctuations\cite{kivelson2008,xms2008}. Neutron scattering
experiments have provided strong evidence showing that the
magnetic fluctuations are indeed nematic above
$T_N$\cite{Harringer2010}. However, the $C^4$ rotational symmetry
breaking can also be caused by other degrees of freedom. First, in
iron-pnictides, the development of the C-AF order is always
accompanied with a lattice structure distortion\cite{La1,Ce1}. A
structural transition temperature $T_s$ from the tetragonal
lattice structure to an orthorhombic lattice structure is either
at or above $T_N$, namely, $T_s\geq T_N$\cite{La1,Ce1}. Second, an
onsite ferro-orbital order between $d_{xz}$ and $d_{yz}$ orbitals
also breaks the same $C^4$ rotational symmetry. Recently, such
ferro-orbital order has been observed by
ARPES\cite{nematic-yi2010,Yi2011-nematic} in pressure de-twinned
samples. From standard symmetry argument, all the orders that
break the same symmetry are allowed to be coupled directly with
each other. Thus the existence of any one of them could lead to
the presence of the others. Therefore, it is difficult to
disentangle the role of the different degrees of freedom behind
the rotational symmetry breaking.

This paper is to review the progress based on the spin-driven
nematic order in Fe-pnictides materials. Before we start our
review, we would like to discuss accumulative evidence supporting
that the spin degree of freedom plays the dominant role in driving
the rotational symmetry breaking. First, the lattice distortion
observed throughout different families of iron-pncitides is rather
small, much less than 1\% of original lattice\cite{La1,Ce1}. Such
a small lattice distortion can not cause a large change in
electronic structures observed in the nematic state. Second,
ferro-orbital ordering observed in
ARPES\cite{nematic-yi2010,Yi2011-nematic,nematic-feng2011} is also
small and a separated ferro-orbital ordering transition is never
observed. More importantly, the change of electronic structure in
nematic state is not dominated by the $d_{xz}$ and $d_{yz}$
orbitals which can cause a $C^4$ rotational symmetry breaking.
Instead, all the orbitals act together and are equally important
in the process of band reconstruction observed in the nematic
state\cite{Hec2010}. Third, even if the ordered magnetic moment in
the C-AF state varies significantly in different families of
iron-pnictides, the total magnetic moment obtained from inelastic
neutron scattering is larger than over $2.0\mu_B$ in underdoped
samples\cite{Wang2011lifeas,Harringer2010}. There is no question
that the fluctuation of such a large moment can cause an
electronic nematicity. Finally, as we will discuss in the paper,
predictions derived from spin-driven mechanism are in a good
agreement with experimental results\cite{xms2008,kivelson2008}.

Iron-pnictides are metallic instead of insulating. However,
electron-electron correlation has been proved to be very
important. Experimentally, it has clearly shown that the
nematicity in iron-pnictides is not a Pomeranchuk type instability
characterized by Fermi surface deformation. Instead, in
iron-pnictides, the entire band structure is reconstructed in the
nematic state\cite{arpes-yang2009,Hec2010}, which suggests that it
is not driven by the states near the Fermi surface, rather by all
electrons due to correlation effects. These two facts lead to
different approaches to study spin-driven nematicity. On one hand,
being a metallic state, it is not surprising to treat the C-AF
state as a spin density wave(SDW) state and formulate the
nematicity based on the two SDW order parameters specified by
order wavevectors $(0,\pi)$ and $(\pi,0)$
respectively\cite{Eremin2011a,nematic-fernandes2010}. On the other
hand, due to the importance of electron-electron correlations as
well as the large magnetic moments observed in neutron scattering,
we can take a local moment to formulate microscopic models based
on short range exchange interactions\cite{xms2008,kivelson2008}.
The nematic transition can be explicitly obtained and the
nematicity can also be reflected in the magnetic exchange
couplings. Fortunately, these two approaches result in an
identical low energy Ginzburg-Laudau effective action.

In this review paper, we will first take a phenomenological
approach. We will discuss the field theory of the coupled nematic
order and various aspects of the system at finite temperature, in
particular, the nature of the phase transitions. Second, we will
analyze microscopic models based on effective magnetic exchange
couplings to obtain quantitative results and discuss the
microscopic parameters that control the nematic transitions.
Finally, we will discuss briefly nematic aspects of
iron-chalcogenides, although at this stage it is still too early
to write a complete review on these materials.

\begin{figure}
\centering
\includegraphics[width=3.0in]{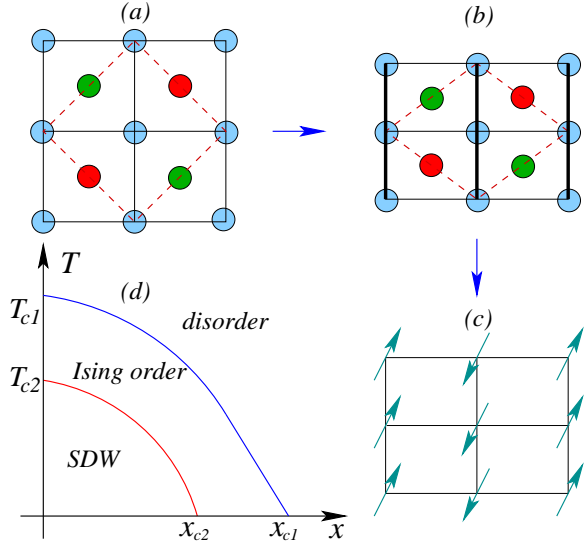}
\caption{$(a)$, the lattice structure of Fe-pnictides materials at
room temperature. The gray circles are Fe ions, the green and red
circles are As ions above and below the Fe plane, respectively.
The dashed square is the two-Fe unit cell. $(b)$, the lattice
structure after the Ising nematic order is developed: the thick
lines represent the bonds between parallelly aligned spins, but an
Ising nematic order does not necessarily imply a nonzero long
range spin order \cite{La1}. $(c)$, the $(0, \pi)$ spin order at
low temperature. $(d)$, the sketchy phase diagram for lattice
distortion and C-AF (or SDW) as a function of temperature and
doping $x$, for some of the Fe-pnictides materials. The blue curve
represents the Ising transition, the red curve represents the C-AF
transition.} \label{phasedia}
\end{figure}

\section{field theory of nematicity}

\subsection{Ginzburg-Landau field theory}

Based on the observation discussed in the previous section, we
first seek a phenomenological field theory description of the
$(\pi, 0)$, $(0, \pi)$ C-AF (or SDW) and the nematic order. It was
first proposed in Ref.~\cite{si2008,Ma2008lu,Yildirim2008a} that
the minimal spin model for the undoped parent Fe-pnictides
material is either a $S = 1$ or $S = 2$ spin model with nearest
and next-nearest neighbor couplings $J_1,J_2$ that depend on the
competition between the onsite Hubbard interaction, crystal field
splitting, and the Hunds rule: \beqn H =
\sum_{<i,j>}J_1\vec{S}_i\cdot\vec{S}_j + \sum_{\ll i,j \gg} J_2
\vec{S}_i\cdot \vec{S}_j. \label{j1j2} \eeqn Unlike cuprates, in
this system $J_2$ is comparable with $J_1$ because the As (or P)
atoms are located at the plaquette center of the square lattice of
magnetic Fe ions, rather than on the links. There is also a much
weaker interlayer coupling $J_{z}$, which is necessary to
stabilize the spin order. It is well-known that when $J_1 < 2J_2$,
the classical ground state manifold of the model in Eq.~\ref{j1j2}
is $S^2 \otimes S^2$, because the two sublattices of the square
lattice will each form a N\'eel order ($\vec{\phi}_1$ and
$\vec{\phi}_2$), and the ground state energy is independent of the
relative angle between these two N\'eel vectors. However, there is
no generic symmetry of the system that protects the degeneracy
within the classical $S^2 \otimes S^2$ ground state manifold
(GSM), thus quantum and thermal fluctuations will both lift the
degeneracy, leading to parallel or antiparallel alignment of the
two sublattice N\'eel vectors
\cite{henley1989,coleman1990,sachdev1991a}, thus the generic GSM
of the system is $S^2 \otimes Z_2$, where the $Z_2$ can be
described by the Ising order parameter $\sigma$ defined as
follows: \beqn \sigma \sim \vec{\phi}_1 \cdot \vec{\phi}_2. \eeqn

We will defer the quantitative discussion of Eq.~\ref{j1j2} to the
next section, right now we will just discuss a Ginzburg-Landau
theory for this lattice model. The lattice structure at high
temperature has a perfect reflection symmetry $\mathrm{P}_{xy}: x
\rightarrow y$, $y \rightarrow x$, and under this transformation,
$\sigma$ always changes sign: $\sigma \rightarrow -\sigma$. Thus
if $\langle \sigma \rangle \neq 0$, the reflection symmetry
P$_{xy}$ has to break, $i.e.$ the system symmetry must reduce from
tetragonal to orthorhombic. $\sigma$ is precisely our nematic
order parameter.

Based on the symmetry of the system, the minimal GL theory
describing the C-AF and nematic order is \cite{xuqi} \beqn F_{GL}
&=& (\nabla_\mu \sigma)^2 + r_{\sigma}\sigma^2 + \sum_{a = 1}^2
(\nabla_\mu\vec{\phi}_a)^2 + r_{\phi}|\vec{\phi}_a|^2 \cr\cr &+&
\tilde{u}\sigma\vec{\phi}_1\cdot\vec{\phi}_2 + \cdots
\label{gl}\eeqn Roughly speaking, $r_\sigma$ and $r_\phi$ are
tuned by the transition temperature for nematic and C-AF order
respectively, large $r$ corresponds to the high temperature
disordered phase. The cubic coupling $\tilde{u}$ in Eq.~\ref{gl}
implies that when $\vec{\phi}_1$ and $\vec{\phi}_2$ are ordered,
the nematic order $\sigma$ has to order; while the reverse is not
necessarily true: there can be a phase with $\sigma$ ordered while
$\vec{\phi}_a$ disordered.

\begin{figure}
\centering
\includegraphics[width=2.5in]{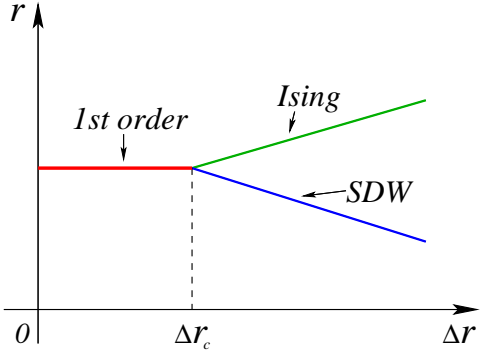}
\caption{The schematic phase diagram of Ginzburg-Landau mean field
theory in Eq. \ref{gl}, plotted against $r = r_\sigma + r_\phi$,
and $\Delta r = r_\phi - r_\sigma$. $r$ is linear with temperature
$T$, while $\Delta r$ is tuned by anisotropy ratio $J_z / J_{in}$.
When $\Delta r$ is small, the interaction between $\vec{\phi}_1$
and $\vec{\phi}_2$ induces a strong first order transition, which
corresponds to the undoped 122 materials with more isotropic
electron kinetics; when $\Delta r$ is large, the transition is
split into two transitions, with an Ising transition followed by
an magnetic transition at lower temperature, and this is the case
in the 1111 materials with quasi two dimensional physics.}
\label{pdmf}
\end{figure}

By simply minimizing the GL theory, we obtain the phase diagram in
Fig.~\ref{pdmf}: When $\Delta r = r_{\phi} - r_{\sigma}$ is large,
meaning $\sigma$ has a much stronger tendency to order compared
with $\vec{\phi}_a$, the system goes through two second order
transitions associated with ordering of $\sigma$ and
$\vec{\phi}_a$ respectively; while when $\Delta r$ is small, the
cubic coupling $\tilde{u}$ in Eq.~\ref{gl} merge the two
transitions into one strong first order transition.

What does $\Delta r$ correspond to in the real system? It is
reasonable to start with a quasi two dimensional structure with
interlayer spin coupling $J_{z}$ much weaker than inplane
couplings $J_1 \sim J_2 \sim J$. In Fe-pnicdides materials, the
inplane coupling $J$ is the energy scale that controls the
ordering temperature of the Ising order parameter $\sigma$.
However, for purely two dimensional systems, the O(3) C-AF is
destroyed by thermal fluctuation at infinitesimal nonzero
temperature, thus there is only one Ising transition at finite
temperature for pure two dimensional system. With small $J_{z}$,
the transition temperature of magnetic order scales as $T_{N} \sim
J/(\log{J / J_{z}}) \ll T_{ising}$. Thus $\Delta r$ is actually
tuned by the effective dimensionality of the system: When the
system is more isotropic, $\Delta r$ is smaller. The phase diagram
of GL theory Eq.~\ref{gl} is shown in Fig.~\ref{pdmf}.

Experiments have shown that in the 1111 materials, at finite
temperature there are two second order transitions associated with
lattice distortion and magnetic transition respectively
\cite{LaFeAs2ndorder}; while in undoped 122 materials there is
only one strong first order transition where both phenomena occur
simultaneously \cite{Ba3,Sr1,Sr2,Sr3,Ca2}. A simple Comparison
between the experimental facts and our GL theory would imply that
the 122 materials are more isotropic than 1111 materials. This is
indeed true in real systems: First of all, the electron band
structure calculated from LDA shows a much weaker $z$ direction
dispersion compared with the 122 samples \cite{122isotropylda};
also the upper critical field $H_{c2}$ of 122 samples is much more
isotropic \cite{isotropic122}. This justifies treating the 1111
materials as a quasi two dimensional system, while treating the
122 materials as a three dimensional one. When $J_z$ and $J$ are
close enough, $\Delta r$ is small, and the coupling between the
nematic order and the C-AF will drive the transition first order
by minimizing the free energy Eq.~\ref{gl}.

\subsection{finite temperature phase transitions}

Now let us focus on the regime with large $\Delta r$, $i.e.$ when
there are two separate second order transitions at finite
temperature. In principle, the transition of the Ising order
parameter $\sigma$ should belong to the 3D Ising Wilson-Fisher
(WF) universality class, while the transition of the C-AF order
parameter $\vec{\phi}$ should belong to the 3d O(3) WF
universality class. Thus at both transitions the specific heat
should have a singularity peak. However, the specific heat
measurement on $\mathrm{BaFe_{2-x}Co_xAs_2}$ reveals two close but
separate transitions at finite temperature, with a sharp peak at
the C-AF transition, and a discontinuity at the lattice distortion
transition \cite{Co1}. A discontinuity of specific heat is a
signature of mean field transition, in contrast to the sharp peak
of Wilson-Fisher fixed point in 3 dimensional space.

Mean field transition can only be accurate for systems with
spatial dimension equal to or higher than 4. Thus the specific
heat measurement in Ref.~\cite{Co1} implies that effectively the
dimension of the nematic transition is enhanced. In this section
we will demonstrate that by coupling to the lattice strain field
fluctuation, the effective spatial dimension $D_{eff}$ for the
nematic transition is enhanced to $D_{eff} = 5$, while the
dimension for C-AF transition remains $D_{eff} = 3$.


Let us first investigate the C-AF transition. The C-AF transition
at finite temperature should belong to the 3D O(3) transition if
the lattice elasticity is ignored. The O(3) order parameter
$\vec{\phi}$ couples to the lattice strain field with a quadratic
term \cite{dh}: \beqn |\vec{\phi}|^2(\partial_x u_x +
\partial_y u_y + \lambda^\prime \partial_z u_z). \eeqn After integrating out the
displacement vector $\vec{u}$, this coupling generates a singular
long range interaction between $|\vec{\phi}|^2$ in the real space:
\beqn \int d^3r d^3r^\prime g|\vec{\phi}|^2_r \frac{f(\vec{r} -
\vec{r}^\prime)}{|r - r^\prime|^3}|\vec{\phi}|^2_{r^\prime}. \eeqn
$f$ is a dimensionless function which depends on the direction of
$\vec{r} - \vec{r}^\prime$. The scaling dimension of $g$ is
$\Delta[g] = \alpha = 2/\nu - 3$, and $\nu$ is the standard
exponent at the 3D O(3) transition, which is greater than $2/3$
according to various types of numerical computations
\cite{vicari2003}. Therefore this long range interaction is
irrelevant at the 3D O(3) transition, and by coupling to the
strain field of the lattice, the C-AF transition is unaffected.
However, if the C-AF has an Ising uniaxial anisotropy, the C-AF
transition becomes a 3D Ising transition with $\nu < 2/3$, and the
strain field would induce a relevant long range interaction
\cite{dh}, which leads to a run-away flow from the 3D Ising fixed
point.

However, since the symmetry of the nematic order parameter
$\sigma$ is the same as the shear strain of the lattice, the
strain tensor will couple to the Ising field $\sigma$ as
\cite{xuqi} \beqn F_{\sigma, \vec{u}} &=& \tilde{\lambda} \sigma
(\partial_x u_y +
\partial_y u_x) + \cdots \eeqn $\vec{u}$ is the displacement
vector, the ellipses are all the elastic modulus terms. Notice
that we have rotated the coordinates by 45 degree, since the true
unit cell of the system is a two Iron unit cell (Fig.~\ref{pdmf}).
After integrating out the displacement vector $\vec{u}$, the
effective free energy of $\sigma$ gains a new singular term at
small momentum: \beqn F_{\theta, \phi} \sim
 f(\theta, \phi)|\sigma_k|^2 . \eeqn $f$ is a function of spherical
 coordinates $\theta$ and $\phi$ defined as
 $(k_x, k_y, k_z) = k (\cos(\theta)\cos(\phi),
\cos(\theta)\sin(\phi), \sin(\theta))$, but $f$ is independent of
the magnitude of momentum $\vec{k}$. By tuning the uniform
susceptibility $r$, at some spherical angle the minima of $f$
reduces to zero, we will call these minima as nodal points. These
nodal points are isolated from each other on the two dimension
unit sphere labelled by $\theta, \ \phi$, and are distributed on
the unit sphere $(\theta, \phi)$ according to the lattice
symmetry.

The symmetry allows the minima to locate at the north and south
pole, $i.e.$ $\theta = 0$ and $\pi$. Expanded at these two minima,
the free energy reads \beqn F = \int q^2 dq \theta d\theta ( q^2 +
\lambda \theta^2 + r)|\sigma_{q, \theta}|^2 + O(\sigma^4) + \cdots
\label{5d}\eeqn The naive power counting shows that effectively
the spatial dimension of this field theory Eq. \ref{5d} is
$D_{eff} =5$. The quartic coupling $\sigma^4$ is irrelevant based
on the straightforward power counting. Therefore the strain tensor
fluctuation effectively increases the dimension by two, which
drives the transition a mean field transition. This result is
completely consistent with the specific heat measurement on
$\mathrm{BaFe_{2-x}Co_xAs_2}$.

Using the Cartesian coordinate, the free energy Eq.~\ref{5d}
becomes: \beqn F = \int dk_x dk_y dk_z (\frac{k_x^2 +
k_y^2}{k_z^2} + k_z^2 + r)|\sigma_k|^2 + \cdots \eeqn Based on
this equation, the effective dimensions for momenta $k_x$, $k_y$
and $k_z$ are \beqn \Delta[k_x] = \Delta[k_y] = 2\Delta[k_z] = 2.
\eeqn Thus the total dimension is still 5, considering
$\Delta[k_x] = \Delta[k_y] = 2\Delta[k_z] = 2$. All the other
momentum dependent terms in this free energy are irrelevant.

Based on the discussions above, we conclude that the lattice
elasticity fluctuation will strongly modify the nature of the
nematic transition, but will leave the C-AF transition unaffected.
In Ref.~\cite{cano2010}, the authors proposed that when the two
transitions are very close, the lattice fluctuation has the
potential of driving the C-AF transition first order, because when
these two transitions are very close the effective quartic term of
the magnetic order parameter can become negative. This effect was
also observed in experiments on lightly electron doped 122
materials in Ref.~\cite{kim2011}.

\subsection{Hertz-Millis theory}

In this subsection we will discuss the quantum critical points
associated with nematic and C-AF orders \cite{xuqi}. A
Hertz-Millis theory for quantum nematic order was studied in
Ref.~\cite{fradkin2001}. However, in our system the nematic order
is strongly correlated with the spin interaction, thus the
Hertz-Millis theory will be very different. The LDA calculation
and photoemission both conclude that in the Brillouin zone of most
Fe-pnictides materials, there are two hole pockets at the $\Gamma$
point ($0, 0$), and two electron pockets located at the M points
($(0, \pi)$ and $(\pi, 0)$). Since the hole and electron pockets
have different shapes, under translation of $\vec{Q} = (\pi, 0)$
in the momentum space, the hole pocket will intersect with the
electron pocket. This intersection leads to overdamping of the SDW
order parameters. The decay rate can be calculated using Fermi's
Golden rule: \beqn \mathrm{Im}[\chi(\omega, q)] &\sim& \int
\frac{d^2k}{(2\pi)^2} [f(\epsilon_{k+q}) - f(\epsilon_{k +
\vec{Q}})] \cr\cr &\times& \delta(\omega - \epsilon_{k+q} +
\epsilon_{k+\vec{Q}}) |\langle k + Q| \vec{\phi}_{i, q} |k+q
\rangle|^2 \ \cr\cr &\sim& \ c_0 \frac{\omega}{|\vec{v}_h \times
\vec{v}_e|}. \label{damp} \eeqn $v_h$ and $v_e$ are the fermi
velocity at the points on the hole and electron pockets which are
connected by wave vector $(\pi, 0)$. The standard Hertz-Millis
\cite{hertz1976} formalism leads to a coupled $z=2$ theory in the
Euclidean momentum space with Lagrangian \beqn L_q &=& \sum_{i =
1}^2 \vec{\phi}_i \cdot (|\omega| + q^2 + r) \vec{\phi}_i  +
\gamma\vec{\phi}_1 (q_x^2 - q_y^2)\cdot \vec{\phi}_2 + L^\prime,
\cr\cr L^\prime &=&
 \tilde{A}(|\vec{\phi}_1|^4 + |\vec{\phi}_2|^4) - \alpha
(\vec{\phi}_1\cdot \vec{\phi}_2)^2 + \tilde{C}
|\vec{\phi}_1|^2|\vec{\phi}_2|^2. \label{field3} \eeqn The
parameter $r$ can be tuned by pressure and doping. The Ising
symmetry of $\sigma = \vec{\phi}_1 \cdot \vec{\phi}_2$ on this
system corresponds to transformation \beqn x &\rightarrow& y, \ \
y \rightarrow x, \cr\cr \vec{\phi}_1 &\rightarrow& \vec{\phi}_1, \
\ \vec{\phi}_2 \rightarrow - \vec{\phi}_2, \ \ \sigma \rightarrow
- \sigma. \label{symm12}\eeqn This Ising symmetry excludes the
term $\vec{\phi}_1 \cdot \vec{\phi}_2$ in the Lagrangian, while
the mixing term $\gamma\vec{\phi}_1 (q_x^2 - q_y^2)\cdot
\vec{\phi}_2$ is allowed.


We can diagonalize the quadratic part of this Lagrangian by
defining $\vec{\phi}_A = (\vec{\phi}_1 + \vec{\phi}_2)/\sqrt{2}$
and $\vec{\phi}_B = (\vec{\phi}_1 - \vec{\phi}_2)/\sqrt{2}$: \beqn
L_q &=& \vec{\phi}_A \cdot (|\omega| + (1 - \frac{\gamma}{2})q^2_x
+ (1+\frac{\gamma}{2})q_y^2 + r) \vec{\phi}_A  \cr\cr &+&
\vec{\phi}_B \cdot (|\omega| + (1+\frac{\gamma}{2}) q^2_x + (1 -
\frac{\gamma}{2})q_y^2 + r) \vec{\phi}_B  + L^\prime, \cr\cr
L^\prime &=&
 A(|\vec{\phi}_A|^4 + |\vec{\phi}_B|^4) +B
(\vec{\phi}_A\cdot \vec{\phi}_B)^2 + C
|\vec{\phi}_A|^2|\vec{\phi}_B|^2. \label{field21}\eeqn After the
redefinition, the Ising transformation becomes \beqn  x
&\rightarrow& y, \ \ y \rightarrow x, \cr\cr \vec{\phi}_A
&\rightarrow& \vec{\phi}_B, \ \ \vec{\phi}_B \rightarrow -
\vec{\phi}_A, \ \ \sigma \rightarrow - \sigma. \label{symmAB}\eeqn
Naively all three quartic terms $A$, $B$ and $C$ are marginal
perturbations on the $z = 2$ mean field theory, a coupled
renormalization group (RG) equation is required to determine the
ultimate fate of these terms. Notice that the anisotropy of the
dispersion of $\vec{\phi}_A$ and $\vec{\phi}_B$ cannot be
eliminated by redefining space and time, therefore the number
$\gamma$ will enter the RG equation as a coefficient. The final
coupled RG equation at the quadratic order for $A$, $B$ and $C$
reads: \beqn \frac{dA}{d\ln l} &=&  - 22A^2 - \frac{1}{2}B^2 -
\frac{3}{2} C^2 - BC, \cr\cr \frac{dB}{d\ln l} &=&  - 5 u B^2 - 8
AB - 8u BC, \cr\cr \frac{dC}{d\ln l} &=&  - u B^2 - 4 AB - 20AC -
4u C^2. \label{rg} \eeqn $u$ is a smooth function of $\gamma$,
which decreases smoothly from $u = 1$ in the isotropic limit with
$\gamma = 0$, to $u = 0$ in the anisotropic limit with $\gamma =
2$. 
The self-energy correction of $\vec{\phi}_a$ from the quartic
terms will lead to the flow of the anisotropy ratio $\gamma$ under
RG, but this flow is at even higher order of the coupling
constants, thus in our current discussion $u$ is just taken a
constant.

\begin{figure}
\centering
\includegraphics[width=2.8in]{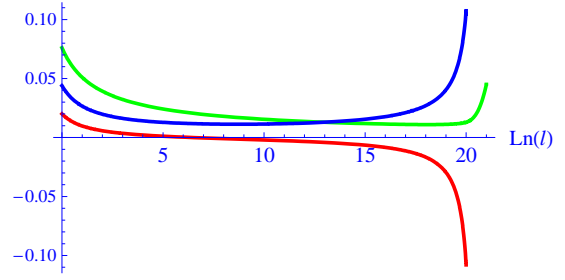}
\caption{The solution of the RG equation Eq. \ref{rg}. All three
quartic perturbations decrease first, then increase and finally
become nonperturbative. The run-away flow is weaker than
marginally relevant perturbations.} \label{rgplot}
\end{figure}

The typical solution of the RG equation Eq.~\ref{rg} is plotted in
Fig.~\ref{rgplot}, for the most natural choice of the initial
values of parameters with $\tilde{A} \gg \alpha$, $\tilde{A} \gg
\tilde{C}$ in Eq. \ref{field3} $i.e.$ the coupling between
$\vec{\phi}_1$ and $\vec{\phi}_2$ is weak. One can see that the
three parameters $A$, $B$ and $C$ all have run-away flows and
eventually become nonperturbative. This run-away flow likely
drives the transition weakly first order. However, the three
coefficients will first decrease and then increase under RG flow.
This behavior implies that this run-away flow is extremely weak,
or more precisely even weaker than marginally relevant
perturbations, because marginally relevant operators will still
monotonically increase under RG flow, although increases slowly.
Therefore in order to see this run-away flow, the correlation
length has to be extremely long $i.e.$ the system has to be very
close to the transition, so the transition remains one single
second order mean field transition for a very large length and
energy range.



The interlayer coupling of $\vec{\phi}_A$ and $\vec{\phi}_B$ has
so far been ignored, this is also a relevant perturbation at the
$z = 2$ mean field fixed point. The $\hat{z}$ direction tunnelling
is written as $J_z \vec{\phi}_{a, z} \cdot \vec{\phi}_{a, z+1}$,
which has scaling dimension 2 at the $z = 2$, $d = 2$ mean field
fixed point, and it becomes nonperturbative when \beqn
\frac{J_{in}}{J_z} \sim (\frac{\xi}{a})^2 \sim r^{-1}. \eeqn This
equation implies that if the tuning parameter $r$ is in the small
window $r < J_z / J_{in}$, the transition crossover back to a $z =
2$, $d = 3$ transition, where all the quartic couplings are
irrelevant. Since in the two dimensional theory these quartic
terms are only very weakly relevant, in the end the interlayer
coupling $J_z$ may win the race of the RG flow, and this
transition becomes one stable mean field second order transition.

\section{Microscopic theory}
\subsection{Model}
As we have mentioned earlier, a simple microscopic model to
capture the C-AF state is the $J_1-J_2-J_z$ Heisenberg model.  The
interlayer coupling $J_z$ must be included in order to address
finite temperature magnetic transition. Moreover, as we will show
later, the layer coupling $J_z$ provides a  control of the
difference  between nematic and magnetic transition temperatures.
The Hamiltonian for such a model can be written as,
\begin{eqnarray} & & H=\sum_n[ J_1\sum_{\langle ij\rangle}\mathbf{S}_{i,n}\cdot\mathbf{S}_{j,n}+J_2\sum_{\langle\langle ij\rangle\rangle}\mathbf{S}_{i,n}\cdot\mathbf{S}_{j,n},\nonumber \\
& & +J_z\sum_{n,i}\mathbf{S}_{i,n}\cdot\mathbf{S}_{i,n+1}],
\label{eq:Hab}\end{eqnarray} where $n$ is layer index.  The spin
nematic order which breaks the $C^4$ rotational symmetry can be
defined as
\begin{eqnarray}
\sigma_i=\frac{1}{2}(\mathbf{S}_{i}\mathbf{S}_{i+\hat
x}-\mathbf{S}_{i}\mathbf{S}_{i+\hat y})
\end{eqnarray}
where $\hat x$ and $\hat y$ are unit vectors in a  two dimensional
tetragonal plane.

As we have mentioned in the introduction that both lattice
distortion and ferro-orbital ordering which break the same  $C^4$
symmetry can be coupled to  the spin nematic order. To be more
specific, we consider the NN coupling
$J_1(|\mathbf{r}_i-\mathbf{r}_j|)$ as a function of the lattice
distance between two sites.  If there is a small orthorhombic
lattice distortion $\delta r$,  the NN coupling becomes \bea
J_1(r_0+\delta r/2)\mathbf{S}_{i}\mathbf{S}_{i+\hat
x}+J_1(r_0-\delta r/2)\mathbf{S}_{i}\mathbf{S}_{i+\hat y}. \eea
where $r_0$ is the lattice constant of the original tetragonal
lattice. We can expand the above term and obtain a coupling
between the lattice distortion and the nematic order,
$\lambda_{sl}(r_0)\delta r\sigma_i$, where
$\lambda_{sl}(r)=\frac{\partial J_1(r)}{\partial r}$.  There is
also coupling between the spin nematic order  and an onsite
orbital order between $d_{xz}$ and $d_{yz}$ orbitals. The coupling
can be written  as \bea \lambda_{so}(n_{xz}(i)-n_{yz}(i))\sigma_i,
\eea where $n_{xz(yz)}(i)$ is electron density at the
corresponding orbitals.  If we include all these couplings and
consider that the magnetism is the dominant player in the system,
we can integrate out the above lattice and orbital fluctuations to
obtain an effective  pure magnetic term to  account for the
couplings.   It is easy to show that such an integration leads to
a biqradratic spin coupling limited to the  NN bonds, namely,
\begin{eqnarray}
H_q=-\sum_{<ij>_{NN},n}K(\mathbf{S}_{i,n}\cdot\mathbf{S}_{j,n})^2.
\end{eqnarray}
By adding such a term into the Hamiltonian in Eq.\ref{eq:Hab}
which will be called as the $J_1-J_2-J_z-K$ model. The C-AF state
is the ground state of the model when $J_1<2J_2$ and $K>0$.  This
extension has been recently considered
in\cite{Wysocki2011,magnetic-hu2011}.

\subsection{Large-S limit}

We can exactly solve above model in the large S-limit. Without the
$K$ term, in the case of $J_1<2J_2$, the ground state is
infinitely degenerate, as the relative angle between the spin
directions on two sublattices is entirely arbitrary. If we
consider  quantum fluctuation,
such a degeneracy will be broken. To show this, we can take Holstein-Primakoff transform for spin operators:\bea& &  S^+_i=\sqrt{2S}a_i\sqrt{1-\frac{a^\dag a_i}{2S}}, S^-_i=\sqrt{2S}a^\dag_i\sqrt{1-\frac{a_i^\dag a_i}{2S}}\nonumber \\
& & S^z_i=S-a^\dag_ia_i,\label{eq:HP_transform}\eea where $a_i$'s
are annihilation operators of spin wave bosons satisfying
$[a_i,a_j]=[a^\dag_i,a^\dag_j]=0$ and
$[a_i,a^\dag_j]=\delta_{ij}$.  We consider the following spin
configuration in two sublattices:
if a site $(i_x,i_y)$ belongs to the first sublattice, its spin moment is $\mathbf{S}_{1,i}=S(0,0,(-1)^{i_x})$; if it belongs to the second sublattice, its spin moment is $\mathbf{S}_{2,i}=S(-1)^{i_x}(\sin\theta,0,\cos\theta)$.  Taking linear spin wave approximation in the Holstein-Primakoff transform, we have \bea S^+_{1,i}&=&\sqrt{2S}(\frac{1+(-1)^{i_x}}{2}a_i-\frac{1-(-1)^{i_x}}{2})a_i^\dag,\\
\nonumber S^-_{1,i}&=&\sqrt{2S}(\frac{1+(-1)^{i_x}}{2}a_i^\dag-\frac{1-(-1)^{i_x}}{2})a_i,\\
\nonumber S^z_{1,i}&=&(-1)^{i_x}(S-a^\dag_{i}a_i),\\
\nonumber
S^+_{2,i}&=&(-1)^{i_x}(\sqrt{\frac{S}{2}}(a_i+a^\dag_i)\cos\theta+(S-a^\dag_ia_i)\sin\theta)\nonumber
\\ & & +\sqrt{\frac{S}{2}}(a_i-a^\dag_i),\\\nonumber
S^-_{2,i}&=&(-1)^{i_x}(\sqrt{\frac{S}{2}}(a_i+a^\dag_i)\cos\theta+(S-a^\dag_ia_i)\sin\theta)\nonumber
\\ & &-\sqrt{\frac{S}{2}}(a_i-a^\dag_i),\\\nonumber
S^z_{2,i}&=&(-1)^{i_x}(-\sqrt{\frac{S}{2}}(a_i+a^\dag_i)\sin\theta+(S-a^\dag_ia_i)\cos\theta).\label{eq:LSWtransform}\eea
Substituting these expressions into Eq.(\ref{eq:Hab}), and taking
Fourier transformation, we obtain \bea
H^{LSW}_=\sum_k[A_ka^\dag_ka_k+\frac{B_k}{2}(a_{k}a_{-k}+h.c.)],\eea
where \bea A_k &=&
4J_1S+4J_2S+2J_1S[(1+\cos\theta)\cos(k_x)\nonumber \\
&+& (1-\cos\theta)\cos(k_y)]+4J_z,\\ \nonumber
B_k&=&-\{2J_1S[(1-\cos\theta)\cos(k_x) \\ &+&
(1+\cos\theta)\cos(k_y)] \nonumber + 2J_2S(\cos(k_x+k_y)
\\ &+& \cos(k_x-k_y))+4J_zcos(k_z)\}.\eea

Taking the Bogliubov transform for bosons, we obtain\bea
H^{LSW}=\sum_k(\gamma^\dag_k\gamma_k+\frac{1}{2})\omega_k,\eea
where \bea\omega_k=\sqrt{A^2_k-B^2_k}, \label{dispersion_sw}\eea
and $ \gamma_k=\cosh\psi_ka_k+\sinh\psi_ka^\dag_{-k},
\gamma^\dag_k=\cosh\psi_ka^\dag_k+\sinh\psi_ka_{-k},$ and $
\tanh(2\psi_k)=-\frac{B_k}{A_k}.$  At zero temperature for small
$J_z$, the quantum fluctuation contributes a zero-point energy for
bosons\bea
E^{Q}_0=\frac{1}{2}\sum_k\omega_k=-\frac{\gamma_QSJ_1^2}{4J_2}\cos^2\theta+J_2O((\frac{J_1}{J_2})^4)\label{eq:OrderByDisorder},\eea
where $\gamma_Q=0.13$.  The ground state, therefore, is either
$\theta=0$ or $\theta=\pi$, which means   that spins in two
sublattice are collinearly aligned, namely, the C-AF state.  The
collinear configuration is an effect of spin fluctuation. This
effect is called ``order by disorder"
\cite{henley1989,coleman1990,sachdev1991a}.

The energy in Eq.\ref{eq:OrderByDisorder} can be taken into
account by the bi-quadratic  $K$ term as well. The quantum
fluctuation contribution to the value of $K$ is given by
$\frac{\gamma_QJ_1^2}{8S^3J_2}$.  We  can thus view the $K$ term
in the $J_1-J_2-J_z-K$ model as an effective term that includes
contributions from all different couplings.   In the C-AF state,
the $K$ term  can be decoupled as \bea
H_{q}=-2\sum_{<ij>_{NN},n}K\sigma_{ij,n}(\mathbf{S}_{i,n}\cdot\mathbf{S}_{j,n})+cons.
\eea where
$\sigma_{ij,n}=<\mathbf{S}_{i,n}\cdot\mathbf{S}_{j,n}>=\pm
\sigma$, is the expectation value of the spin-spin interaction in
the C-AF state.  The sign is chosen to be positive if the spins
are FM along one axis (b-axis) and negative if  they are AF along
the other axis (a-axis) in the C-AF state. Combining this term
with $J_1$ term in the original Hamiltonain, the effective nearest
neighbour exchange couplings in the C-AF State become \bea
J_{1a}=J_1+ 2K\sigma, J_{1b}=J_1-2K\sigma. \eea Thus, the spin
wave excitations in the C-AF state are effectively described by a
$J_{1a}-J_{1b}-J_2-J_z$ model.

\begin{figure}[t]
\centering
\includegraphics[width=2.8in]{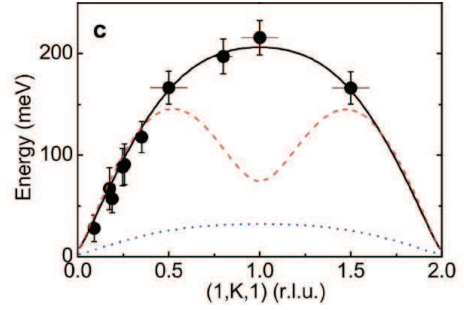}
\caption{ The spin wave dispersion along $(\Pi, k, \Pi)$ direction
reported in\cite{Zhaojun2009}: the solid line is fitted when
$J_{1a}>>J_{1b}$ and the dashed line is fitted if $J_{1a}\sim J
_{1b}$.}\label{spinwave122}
\end{figure}

Experimentally, a large difference between $J_{1a}$ and $J_{1b}$
is indeed observed in the C-AF
state\cite{Zhao2008dai,Zhaojun2009,Harringer2010}.  The spin wave
energy at wavevector  $(\pi,\pi,\pi)$ is very sensitive to the
difference between $J_{1a}$ and $J_{1b}$. The energy is
proportional to $\sqrt{J_{1a}-J_{1b}}$. Experimentally, The gap
$\omega(\pi,\pi,\pi)$ measured  in the 122 family has been very
large as shown in fig.\ref{spinwave122}, namely, the difference of
$J_{1a}$ and $J_{1b}$ is very large. Such a large difference
suggests a strong nematic order and a sizable $K$ value.

\subsection{large-N limit}

\begin{figure}[t]
\centering
\includegraphics[width=8cm]{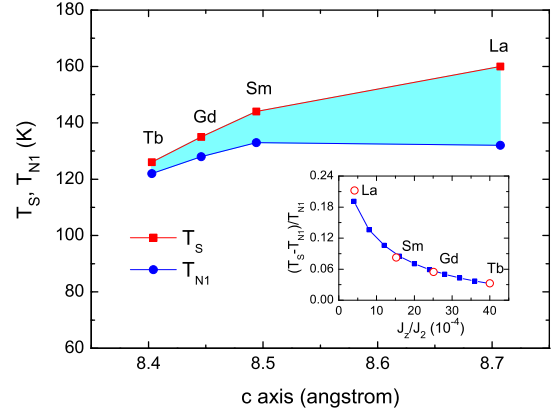}
\caption{Plot of structural phase transition temperature $T_s$,
and magnetic transition temperature $T_{N}$ associated to the AF
order of Fe ions, versus the c-axis for RFeAsO (R = La, Sm, Gd,
Tb). Inset: plot of $(T_S-T_{N})/T_{N}$ versus $J_z/J_2$. The open
circles denote the experimental data of $(T_S-T_{N})/T_{N}$, and
the solid squares denote the theoretical
values\cite{Louyk2009a}.}\label{fig:TsAndTn_exp}
\end{figure}

The model can be analytically solved in the large-N limit to
obtain  phase diagram and transition temperature.  We take  the
continuum limit following the derivation in \cite{
Fagn2008nematic,magnetic-hu2011}. The low energy effective
Hamiltonian reads\bea\label{eq:Heff}  H_{eff} =
\int{d}^2r\sum_{n,\alpha}[\tilde{J}_2|\nabla\mathbf{n}_{\alpha,n}|^2
-\tilde{J}_z\mathbf{n}_{\alpha,n}\cdot\mathbf{n}_{n+1,\alpha}]\nonumber \\
- \tilde{K}\sum_n[\mathbf{n}_{1,n}\cdot\mathbf{n}_{2,n}]^2
+\frac{\tilde{J}_1}{2}\sum_n\mathbf{n}_{1,n}\cdot\partial_x\partial_y\mathbf{n}_{2,n}
\eea where $\tilde{J}_i=J_iS^2$, $\tilde{K}=KS^4/2$  and
$\mathbf{n}_{1(2),n}$ are AF orders in the two sublattices
mentioned earlier respectively  and the nematic order in the
continuum limit becomes
$\sigma=\mathbf{n}_{1,n}\cdot\mathbf{n}_{2,n}$. We can solve the
effective model in the large-$N$ limit analytically. In general,
there are two transition temperatures.  The nematic transition
temperature $T_{s}$ is analytically given by
\bea\label{eq:nem_temp}\frac{4\pi\tilde{J}_2}{NT_{s}}=\ln\frac{\tilde{J}_2/(NT_{s})}{\sqrt{(\frac{\tilde{K}}{4\pi\tilde{J}_2})^2+(\frac{\tilde{J}_z}{NT_{s}})^2}+\frac{\tilde{K}}{4\pi\tilde{J}_2}},\eea
\begin{figure}[t]
\centering
\includegraphics[width=8cm]{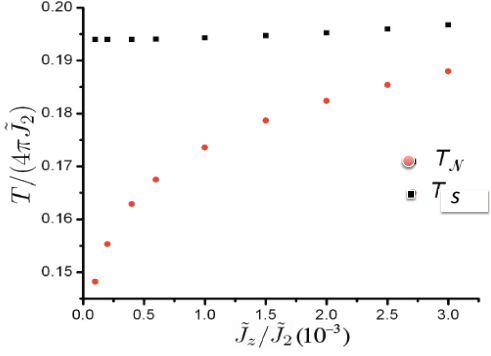}
\includegraphics[width=8cm]{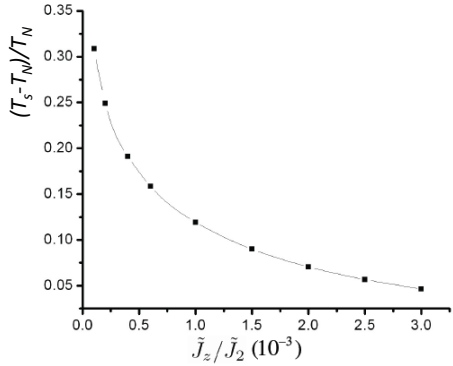}
\caption{Left:$T_{s}$ and $T_{N}$ as the function of $\tilde J_z$
for $\tilde J_2=2\tilde J_1$ and $N=3$. Right:
$\frac{T_{s}-T_{N}}{T_{N}}$ as the function of $\tilde J_z$ for
$\tilde J_1=0.5\tilde J_2$.}\label{fig:TsAndTn}
\end{figure}
and  the C-AF  transition temperatures, $T_{N}$ is determined by,
\bea\label{eq:SDW_temp}& & \frac{\sigma_c}{2\tilde{K}}=\frac{1}{8\pi\tilde{J}_2}\ln\frac{\sigma_c+\frac{\tilde{J}_z}{NT_{N}}+2\sqrt{\frac{\sigma_c^2}{4}+\frac{\sigma_c\tilde{J}_z}{NT_{N}}}}{\tilde{J}_z/(NT_{N})}, \nonumber \\
& &
\frac{\sigma_{c}}{2\tilde{K}}+\frac{1}{NT_{N}}=\frac{1}{4\pi\tilde{J}_2}\ln\frac{\tilde{J}_2}{\tilde{J}_z}.\eea
$T_{N}$ is always lower than $T_{s}$ in the solutions of above
equations. Moreover,  $T_{N}=0$ if $J_z=0$, and is  finite
otherwise. For small $J_z$, one can approximately have
$T_{N}\sim\tilde{J}_2/\ln(\frac{\tilde{J}_z}{\tilde{J}_z})$.  In
fig.\ref{fig:TsAndTn}, we plot the two transition temperatures and
their difference as the function of $J_z/J_2$.  The difference
decreases as $J_z$ increases. This qualitative prediction is
consistent with the GL field theory approach discussed in the
previous section, and it is also confirmed in the 1111
family\cite{Louyk2009a} as shown in fig.\ref{fig:TsAndTn_exp}.

This calculation shows that as soon as $J_z$ is much larger than
$10^{-3}J_z$, the two transitions are close to each other so that
they become practically inseparable in experiments. This result is
consistent with the fact there is only one phase transition in the
parent compounds of the 122 families where $J_z$ determined by
spin wave excitations is rather large: in $\mathrm{CaFe_2Se_2}$,
$J_z$ is almost one third of the in-plane coupling
$J_2$\cite{Zhaojun2009}. In $\mathrm{BaFe_2As_2}$, $ J_z$  is
about $0.015J_2$\cite{Harringer2010}. The prediction is further
confirmed in under-doped 122 materials. Upon electron doping in
$\mathrm{BaFe_2As_2}$, for example,
$\mathrm{BaFe_{2-x}Ni_xAs_2}$\cite{Harriger2009}, the spin
excitations become much less three dimensional. The effective
$J_z$ is drastically reduced and two phase transitions were
observed.


\section{Nematic order in Iron-Chalcogenides}

Iron-chalcogenides include two families, the 11 family (FeTe/Se)
and the 122 family ($\mathrm{A_{1-x}Fe_{2-y}Se_2}$
$(\mathrm{A=K,Rb, Cs})$. While the former shares similar band
structures with iron-pnictides, the latter exhibits several
distinct characters that are noticeably absent in other iron-based
superconductors, including the absence of hole pockets at $\Gamma$
point of Brillouin zone in their superconducting (SC)
phases\cite{Zhangy2011a,WangXP2011,Mou2011}, AF ordered insulating
phases\cite{Baowei2011a, Baowei2011b,spinwave-wang2011a} with very
high N\'{e}el transition temperatures in their parental
compounds\cite{Liu2011a} and intrinsic vacancy ordering. These
intriguing properties have stimulated many exciting research
activities.

\begin{figure}[t]
\includegraphics[width=3.1in]{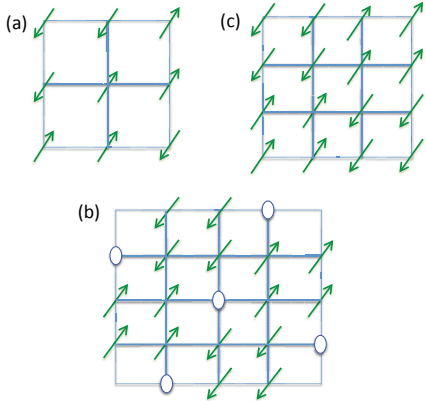}
\caption{Sketch of spin configurations of  (a) bi-collinear AF
(B-CAF) state observed in FeTe, (b) block AF with vacancy order
(B-AF$_{v} $) observed in $\mathrm{K_{0.8}Fe_{1.6}Se_2}$, and (c)
block AF order (B-AF) order. }\label{bcaf}
\end{figure}

While it is too early to call a reasonable review for
iron-chalcogenides, the magnetic orders in the parent compounds of
these two families have been measured. Unlike iron-pnictides whose
different families share a common C-AF state, different families
of iron-chalcogenides exhibit different magnetically ordered
states. The FeTe compounds of the 11 family has a bicollinear
AF(B-CAF) ground state as shown in fig.\ref{bcaf}(a). The B-CAF
state is accompanied by a monoclinic lattice distortion. The
distortion can produce 1.3\% change of the tetragonal lattice
constant, a value much larger than the orthorhombic distortion in
iron-pnictides\cite{Lisl2009fete}. The large lattice distortion
indicates a strong coupling between spin and lattice and is
consistent with the fact there is only one strong first-order type
transition in FeTe, namely, the B-CAF and monoclinic lattice
distortion take place at the same transition temperature which is
about 70K\cite{Chenf2009a}. The ordered magnetic moment in
$\mathrm{FeTe}$ about $2.3\mu_B$ is also significantly larger than
the ones in iron-pnictides.

For the 122 family, $\mathrm{R_{1-x}}$Fe$_{2-y}$Se$_2$, one
magnetically ordered state with iron vacancy order, called
block-AF$_v$ has been identified when $x=0.8$ and $y=0.4$ as shown
in fig.\ref{bcaf}(b). Iron vacancies form a $\sqrt{5}\times
\sqrt{5}$ pattern and four iron spins group together to form a
super-site with ferromagnetic(FM) alignment. In the lattice formed
by the super-sites, it is AF ordered. The ordered moment per iron
site in the B-AF$_v$ state reaches 3.3$\mu_B$ and the transition
temperature is about 500k\cite{Baowei2011a, Baowei2011b}. The
lattice distortion in the B-AF$V$ state can reach 2\% of the
original lattice constant. More importantly, the B-AF$_v$ state is
insulating with a gap larger than 0.3$ev$. It is the first
insulating state observed in iron-based superconductors.

Recently, it has been shown that the effective magnetic exchange
models that describe different ordered states can be unified into
a $J_1-J_2-J_3-K-J_z$
model\cite{Fang2009fete,vacancy-fang2011,magnetic-hu2011}, where
$J_3$  is the next next nearest neighbour magnetic exchange
coupling. Compared to the $J_1-J_2-K-J_z$ model for
iron-pnictides, there are two major differences: $J_1$ is strong
ferromagnetic (FM) in iron chalcogenides  and  $J_3$ is
significant and AF.  Without vacancy, the model has two degenerate
classical ground states: the B-CAF or bloch-AF (B-AF) state which
is shown in fig.\ref{bcaf}(c). An effective theory based on this
model has been discussed in \cite{Xuhu2009}. The complete field
description of the $J_1-J_2-J_3-J_c-K$ model near the B-CAF or
B-AF phase is given by $H_{T}= H_{B}+H_{K}$ where
\begin{eqnarray}
H_{B} =  \int d^2{\bf r}\sum_{n,\alpha}\Big[\frac 1 2 J_3|\nabla
{\vec \phi}_{n,\alpha}({\bf r})|^2 -  J_c \vec\phi_{n,\alpha}({\bf
r})\cdot \vec \phi_{n+1,\alpha}({\bf r})\Big] \nonumber \\ -
g'\sum_{n}\{\left[\vec \phi_{n,1}({\bf r})\cdot {\vec
\phi}_{n,3}({\bf r})\right]^2+\left[\vec \phi_{n,2}({\bf r})\cdot
{\vec \phi}_{n,4}({\bf r})\right]^2\},
\end{eqnarray}
and \begin{eqnarray} H_{K} = - K \int d^2{\bf
r}\sum_{n}\{\left[\vec \phi_{n,1}({\bf r})\cdot {\vec
\phi}_{n,2}({\bf r})\right]^2 \\ + \left[\vec \phi_{n,1}({\bf
r})\cdot {\vec \phi}_{n,4}({\bf r})\right]^2 \nonumber \\ +
\left[\vec \phi_{n,2}({\bf r})\cdot {\vec \phi}_{n,3}({\bf
r})\right]^2 +\left[\vec \phi_{n,3}({\bf r})\cdot {\vec
\phi}_{n,4}({\bf r})\right]^2\}\end{eqnarray} where $\vec
\phi_{n,\alpha}$ with $\alpha=1,2,3,4$ are four Neel order
parameters defined in the four sublattices of the tretragonal
lattice specified by  the $J_3$ coupling  and  $g' \sim 0.13
J_2^2/  J_3$. In this model, there are four independent Ising
orders $\sigma_{1}=g'\vec \phi_{n,1}({\bf r})\cdot {\vec
\phi}_{n,3}({\bf r})$, $\sigma_{2}=g'\vec \phi_{n,2}({\bf r})\cdot
{\vec \phi}_{n,4}({\bf r})$, $\sigma_{+}=K(\vec \phi_{n,1}({\bf
r})+  {\vec \phi}_{n,3}({\bf r}))\cdot (\vec \phi_{n,2}({\bf r})+
{\vec \phi}_{n,4}({\bf r}))$ and  $\sigma_{-}=K(\vec
\phi_{n,1}({\bf r})-  {\vec \phi}_{n,3}({\bf r}))\cdot (\vec
\phi_{n,2}({\bf r})+ {\vec \phi}_{n,4}({\bf r}))$.  $\sigma_{+}$
and $\sigma_{-}$ describe the Ising orders associated with the
B-AF and B-CAF phases respectively.  The Ising  order $\sigma_{1}$
breaks rotational symmetry and represents a nematic phase. The
analysis of this effective field theory has been carried out
in\cite{Xuhu2009,magnetic-hu2011}.  This effective theory also
suggests that the lattice-spin coupling is extremely important in
determining the true magnetic order in iron-chalcogenides due to
intrinsic  magnetic frustration.

Since there are few experiments that have been implemented to
probe the nematic nature of the B-CAF state explicitly, we will
not extend our discussion further here. However, it is interesting
to point out that since the 122 family of iron-chalcogenides may
approach a magnetically ordered insulating phase, the materials
may carry similar nematic phases as curpates upon doping. The
charge degree of freedom hence may manifest itself  and play an
important role in driving nematicity in these new materials.

\section{Summary and Extensions}

In this article we reviewed the study of the electronic nematic
order and the C-AF order that have been universally observed in
iron-pnictides materials. Based on the experimental observation of
the close relation between these two order parameters, we propose
the theory of spin interaction driven nematic order in these
materials. In this article we have thoroughly studied this problem
with both low energy field theories such as Ginzburg-Landau theory
and Hertz-Millis theory, and also microscopic theories in the
large$-S$ and large$-N$ limit. We also briefly discussed the
generalization of this picture to iron-chalcogenides materials.

A proof of the nematic order induced by spin interactions can have
strong impact on the field. it would automatically suggest that
the superconductivity mechanism is most likely to be magnetically
driven. While both experiments and theories reviewed here have
provided convincing evidence, we agree that the debate over the
origin of the rotational symmetry breaking has not been fully
settled at this stage.

Although we have a detailed theory of the interplay of electron
nematicity and spin order, the relation between these order
parameters and the high temperature superconductor is still
unclear. Strong competition between spin orders and superconductor
phase has been observed experimentally in iron-pnictides, meaning
that a complete description of low energy physics should also
involve the superconducting phase. This is one of the directions
that can be pursued in the future.

In section II, we have seen that the lattice strain field
fluctuation strongly affects the finite temperature phase
transition of the nematic order. However, in section II the
discussion of the lattice strain field was purely classical and
phenomenological. At zero temperature, the nematic order will
conceivably strongly couple to the phonon modes of the lattice,
thus the phonon fluctuation will most likely also strongly affect
the quantum critical behavior, as long as the spin wave excitation
at zero temperature. A complete theory including the phonon of the
lattice will significantly enrich the physics.

The 122 family of iron-chalcogenides has opened a new chapter in
the field. It provides a fresh ground to test the concept of
nematism. Since electron-electron correlation is generally
believed to be stronger in iron-chalcogenides than iron-pnictides,
exciting results related to nematism in the future can be
expected.

{\it Acknowledge:} We thank S. Kivelson, S. Sachdev,  C. Fang,
Pengcheng Dai, Donglei Feng and X.H. Chen for many valuable
discussions.









\end{document}